\documentstyle[12pt]{article}
\newcommand{\be}{\begin{equation}}
\newcommand{\ee}{\end{equation}}
\newcommand{\bea}{\begin{eqnarray}}
\newcommand{\eea}{\end{eqnarray}}
\newcommand{\beas}{\begin{eqnarray*}}
\newcommand{\eeas}{\end{eqnarray*}}
\newcommand{\bi}{\begin{itemize}}
\newcommand{\ei}{\end{itemize}}
\newcommand{\bc}{\begin{center}}
\newcommand{\ec}{\end{center}}
\newcommand{\bfl}{\begin{flushleft}}
\newcommand{\efl}{\end{flushleft}}
\newcommand{\bfr}{\begin{flushright}}
\newcommand{\efr}{\end{flushright}}
\newcommand{\f}{\frac}

\def\6{\partial} \def\a{\alpha} \def\b{\beta}
 \def\d{\delta}

  \def\l{\lambda}
\def\m{\mu} \def\n{\nu}

\def\ud#1#2{^{#1}\,_{#2}}
\def\du#1#2{_{#1}\,^{#2}}

\newcommand\I{\leavevmode\hbox{\small1\kern-3.8pt\normalsize1}}
\newcommand\pa{\partial}
\newcommand\bq{\begin{eqnarray}}
\newcommand\eq{\end{eqnarray}}
\newcommand\bes{\begin{equation*}}
\newcommand\ees{\end{equation*}}
\newcommand\bqs{\begin{eqnarray*}}
\newcommand\eqs{\end{eqnarray*}}
\newcommand\nn{\nonumber}

\begin{document}

\title{Symplectic Projector and Physical Degrees of Freedom of The 
Classical Particle}

\author{M. A. De Andrade$^{a}$\footnote{E-mail: marco@cbpf.br}, 
M. A. Santos$^{b}$\footnote{Email:masantos@ufrrj.br} and I. V. Vancea$^{c}$
\footnote{E-mail: ion@dfm.ffclrp.usp.br, ivancea@ift.unesp.br.}}

\date{\small $^a$Grupo de F\'{\i}sica Te\'{o}rica, Universidade Cat\'{o}lica de Petr\'{o}polis \\ 
Rua Bar\~ao de Amazonas 124, 25685-070,  Petr\'opolis - RJ, Brasil \\
\vspace{.5cm}
$^b$Departamento de F\'{\i}sica, Universidade Federal Rural do Rio de Janeiro\\
23851-180, Serop\'{e}dica - RJ, Brasil\\
\vspace{.5cm} 
$^c$
Departamento de F\'{\i}sica e Matem\'{a}tica,\\ Faculdade de Filosofia, Ci\^{e}ncias e Letras de Ribeir\~{a}o Preto
USP,\\ Av. Bandeirantes 3900, Ribeir\~{a}o Preto 14040-901, SP, Brasil
}

\maketitle

\abstract{
The symplectic projector method is applied to derive the local 
physical degrees of freedom of a classical particle moving freely on an 
arbitrary surface. The dependence of the projector on the coordinates and momenta of the particle is discussed.}

\newpage

\section{Introduction}

The problem of finding the physical degrees of freedom of a constrained system
can be traced back to the work of Dirac \cite{Dirac} and represents the 
central issue in the quantization of realistic models. The classical method to 
deal with this sort of problems is to enlarge the phase space by adding 
non-physical variables to the original ones and to define the physical surface
with the help of a nilpotent operator acting on the all degrees of freedom. 
This is the well known BRST method (see \cite{BRST} for a reveiw.) However, in
many cases when the structure of the constraints is simpler, there are other 
ways of finding the physical degrees of freedom. In particular, when the 
constraints are of second class only, a local symplectic projector can be 
written down and the local physical degrees of freedom can be computed from it
(\cite{Amaral1,Amaral2,Marco1,Marco2,Marco3,Marco4,Marco5,Marco6,marcotese}). 
A class of systems that may exhibit simple second class constraints is given by a particle moving on a surface defined by a holonomic function $f(x)=0$. The way in which such system can arise and the problems related to the quantization of it were addressed in \cite{jaffe}. In this letter we are going to study the dynamics of the classical particle on a surface from the point of view of the symplectic projector method. This study is usefull for understanding the local degrees of freedom of the particle as well as the range of the applicability of the method to constrained systems. The paper is organized as follows. In Section 2 we review the method of the symplectic projector. In  Section 3 we derive the local degrees of freedom for a free particle on a surface embedded in $R^n$ and the corresponding Hamiltonian. The last section is devoted to discussions.

\section{Brief Review of The Symplectic Projector Method}

The symplectic projector can be defined for any system subject to second class constraints 
$\phi^m \left(\xi^M \right)=0$. Here $\xi^M = (x^a, p_a )$, $M=1,2,\ldots,2N$
are the coordinates in the phase space which is assumed to be isomorphic to $R^{2N}$ and $m=1,2,\ldots, r=2k$. The symplectic projector has the following form \cite{Amaral2}
\begin{equation}
\Lambda ^{MN} = \;\; \delta ^{\,MN}- J^{ML}\,\frac{\delta{\phi}_{m}}{\delta
\xi ^{L}}\,\Delta^{-1}_{mn}\, \frac{\delta \phi _{n}} {\delta \xi^{N}},
\label{m7}
\end{equation}
where $J^{MN}$ is the symplectic two-form and $\Delta^{-1}_{mn}$ is the inverse of the matrix
\begin{equation}
\Delta_{mn} = \{\phi_m, \phi_n \}.  \label{submatrix}
\end{equation}
The symplectic projector given by the relation (\ref{m7}) projects the phase space variables $\xi^{M}$ onto local variables on the constraint surface 
$\mathbf{\xi }^{*}$ 
\begin{equation}
\xi ^{*M}=\Lambda ^{MN}\xi ^{N}.  \label{m8}
\end{equation}
Starting with a $2N$-dimensional phase-space and $2k$ second class
constraints, we are led to a vector with $2(N-k)$ independent
components. As was noted in \cite{Amaral1}, the boundary conditions that should be satisfied by the normal coordinates and the corresponding momenta to the constraint surface are given by the following relations
\bea
x^{m}(t) &=& \phi^{m}(x)=0,  \label{ion-bc}\\
p_m(t)&=&{\dot{p}}_m(t)=0.  \label{ion-bcmom}
\eea
Then, the physical Hamiltonian
is given by the original one written in terms of the coordinates that are
obtained after projection, i. e. the local coordinates on the physical
surface. These variables are independent, free of constraints and they obey
canonical commutation relations. The equations of motion follow from the
usual Hamilton-Jacobi equations: 
\begin{equation}
\stackrel{.}{\mathbf{\xi }}^{*}=\left\{ \mathbf{\xi }^{*},H^{*}\right\},
\label{m9}
\end{equation}
where $\{~,~ \}$ are the Poisson brackets. From the analogy between the Dirac matrix
\begin{equation}
D^{MN}=\{\xi ^{M}\;,\;\xi ^{N}\}_{D}=J^{MN}-J^{ML}J^{KN}\,\frac{\delta \phi
_{m}}{\delta \xi ^{L}}\,\Delta _{mn}^{-1}\,\frac{\delta \phi _{n}}{\delta
\xi ^{K}},  \label{m10}
\end{equation}
and the projector given by the relation (\ref{m7}) one can see \cite{Marco6} that the following relation holds: 
\begin{equation}
\Lambda =-DJ.  \label{m11}
\end{equation}
We note that the trace of the projector matrix gives the degrees of freedom of the system. In order to quantize the theory, one should start with the physical Hamiltonian obtained above. Then, the observables of the quantum theory should depend only on the coordinates 
$\mathbf{\xi }^*$.

\section{Free Particle on a Surface $f(x)=0$ in $R^N$}

Let us consider a particle moving freely on a smooth arbitrary surface $\Sigma$ 
embedded in $R^{N}$ and defined analytically by the equation
\be
\Sigma:~f(x)=0,
\label{constrsurf}
\ee
where 
$x = \{ x^a \}$, $a=1,2,\cdots,N$ are the Cartezian coordinates in 
$R^N$. The movement on the surface $\Sigma$ can be obtained by introducing the $f\left( x \right)$ in the Hamiltonian through a Lagrange multiplier 
$\lambda$
\be
H=\frac{1}{2m}p_ap^a-\lambda f(x)
\label{Hamiltonian}
\ee
and then interpreting $\lambda$ as an independent variable. Thus, the extended phase space
is coordinatised by $x^\mu=(\lambda,x^a)$ and $p_\mu=(p_\lambda,p_a)$ and it
is endowed with an Euclidean metric and a symplectic two-form. The constraints
on the dynamics are derived in the usual manner, by the Dirac algorithm which
gives the following set of equations
\bea
\phi_1 &=& p_\lambda
\label{constraint1}\\
\phi_2 &=& f(x)
\label{constraint2}\\
\phi_3 &=& \frac{p^a}{m}\frac{\pa f}{\pa x^a}
\label{constraint3}\\
\phi_4&=&\frac{p^a\,p^b}{m^2}\,\pa_a\pa_bf+\frac{\lambda}{m}\,\pa^af\pa_af\
\label{constraint4}.
\eea
The constraints (\ref{constraint1})-(\ref{constraint4}) are of second class,
therefore one can apply the Symplectic Projector Method to determine the 
local physical degrees of freedom \cite{Amaral1,Marco6}. To this end, one 
computes firstly the Dirac brackets in the extended phase space. The non-zero
Dirac brackets are 
\bea
\{\lambda~,~x^b\}_D &=& k^b\nonumber\\
\{\lambda~,~p_b\}_D &= &v_b\nonumber\\
\{x^a~,~p_b\}_D &=&\delta\ud ab-n^an_b=\widetilde\delta\ud ab\nonumber\\
\{p_a~,~p_b\}_D &=&\omega_{ab}\label{Diracbrackets},
\eq 
where we have used the following notations
\bea
k^b&=&\frac{2p^c}{m|\pa f|}\,\pa_cn^b  \label{k}\\
v_b&=&-\frac{p^cp^d}{m|\pa f|}\,\left[ \pa_c\pa_dn_b+3(\pa_cn^a)(\pa_dn_a)\,n_b\right]-2\lambda\,n^c\,\pa_cn_b \label{v}\\
\omega_{ab}&=&p^c(n_b\pa_cn_a-n_a\pa_cn_b)=-\omega_{ba}\label{omega}\\
\eea
The normal vector to the surface $n^a$ has unit norm and is defined in the 
usual way
\be
n_a=\frac {1}{\sqrt{m\alpha}}\,\pa_af=\frac{\pa_af}{|\pa f|}~~~,~~~|\pa f|
\equiv\left|\sqrt{\pa^af\pa_af}\right|\label{normalvector}.
\ee
If we write the components of a symplectic vector in the extended phase space in the order
$(\lambda, x^a, p_\lambda, p_a)$ then the Dirac brackets can be disposed into 
a $(2N+1) \times (2N+1)$ matrix which is called the Dirac matrix 
\be
D= \left( 
\begin{array}{cccc}
0 & k^b & 0 & v_b \\
-k^a & 0 & 0 & \widetilde\delta\ud ab  \\
0 & 0 & 0 & 0  \\
-v_a & -\widetilde\delta\du ab  & 0 & \omega_{ab}
\end{array}
\right).\label{Diracmatrix}
\ee
The symplectic projector is given by the equation (\ref{m11})  and has the following form
\be
\Lambda= \left( 
\begin{array}{cccc}
0 & v_b & 0 & -k^b \\
0 & \widetilde\delta\ud ab & k^a & 0  \\
0 & 0 & 0 & 0  \\
0 & \omega_{ab} & v_ a & \widetilde\delta\du ab
\end{array}
\right).\label{Projector}
\ee
By acting with the projector (\ref{Projector}) on the degrees of freedom of the
extended phase space we obtain the local physical degrees of freedom of the
particle \cite{Amaral1}. Locality, in this case, has the meaning of ``local 
on the surface $\Sigma$''. Note that the trace of the matrix (\ref{Projector})
is equal to the number of the degrees of freedom, which in this case is 
$2N-2$. The projected degrees of freedom are given by the following relations
\bq
\lambda^\ast&=&v_bx^b-k^bp_b\nn\\
x^{\ast a}&=&\widetilde\delta\ud abx^b+k^ap_\lambda\nn\\
p^\ast_\lambda&=&0\nn\\
p^\ast_a&=&\omega_{ab}x^b+v_ap_\lambda+\widetilde\delta\du abp_b\;.
\label{Projectedcoordinates}
\eq
Now let us discuss the above sytem. We can see that there are some distinct cases to deal with.

\subsection{General $\Lambda \left( \xi \right)$}

In the case of an arbitrary function $f$, the entries of the projector matrix are functions of the coordinates of the particle in the phase space. Therefore, there is no general method to separate the linearly independent physical coordinates $\xi^{*\m}$, where $\m, \n \ldots = 1,2,\ldots 2(N-1)$ from the linearly dependent ones $\xi^{*\a}$, where $\a, \b, \ldots = 1,2,3,4$. However, we can derive some general relations to be satisfied by the linearly dependent coordinates as well as by the projector. Since there are constants $c^{\a}_{\m}$ such that 
\be
\xi^{*\a} = c^{\a}_{\m}\xi^{*\m},
\label{lindependence}
\ee
then one can use the relation $\Lambda^2= \Lambda$ to find the following relations that should be satisfied by $c^{\a}_{\m}$'s
\bea
\left( \d^{\m}_{\n} - \Lambda^{\m}_{\a}\left(\xi \right)c^{\a}_{\n}\right)\Lambda^{\n}_{N}\left( \xi \right)\xi^N &=& 0,
\label{const1}\\
\left( \d^{\a}_{\b} - \Lambda^{\a}_{\b}\left(\xi \right)\right)c^{\b}_{\m}\Lambda^{\m}_N\left(\xi \right)\xi^N &=& 0.
\label{const2}
\eea
The above equations should be satisfied simultaneously for all $\n$ and all $\b$. By comparing the number of equations with the number of constants $c$'s, we see that $c$'s may be completely determined only if $N=1,2$. This condition, although necessary, is not sufficient. Indeed, if we treat  (\ref{const1}) and (\ref{const2}) as a system that solves $c$'s, we see that the constants $c^{\a}_{\m}$ are expressed as functions on $\xi$'s. However, by hypothesis, $c$'s should be real numbers which implies that these functions be constants.

Let us look at the time evolution of an arbitrary projected coordinate $\xi^{*M}$. On the constraint surface, it is given by its Poisson bracket with the physical Hamiltonian $H^{*}$
\be
\dot{\xi}^{*M} = \{ \xi^{*M}, H^* \}^{*}_{PB},
\label{starevol}
\ee
where $H^* = H \left( \xi^* \right)$ depends only on the linearly independent variables and $*$ in the Poisson brackets means that they should be computed using the projected variables. In particular, the relation (\ref{starevol}) guarantees that the mapping by $\Lambda$ is canonical.
On the other hand-side, if we interpret $\xi^{*M}\left(\xi\right)$ as function on the phase space, then its time evolution should be given by the following relation
\be
\dot{\xi}^{*M}\left( \xi \right) = \{ \xi^{*M}\left( \xi \right), H \left( \xi \right) \}_{DB}.
\label{normalevol}
\ee
However, on the constraint surface $\Sigma$, the relations (\ref{starevol}) and (\ref{normalevol}) should coincide, i. e. the following relation should hold
\be
\{ \xi^{*M}, H^* \left(\xi^* \right) \}^{*}_{PB} =
\{ \Lambda^{M}_{N}\left( \xi \right) \xi^{N}, H \left(\xi \right)\}_{DB}\left|_{\Sigma = \{ \phi^m \left( \xi \right) = 0 \}}\right.
\label{constrlambda}
\ee
The above relation represents a consistency condition that should be satisfied by the symplectic projector.

\subsection{Constant Symplectic Projector }

There are surfaces for which the entries of the symplectic projector are constant functions on the coordinates $\xi$. These surfaces will satisfy the following relations
\bea
&&\frac{2p^c}{m|\pa f|}\,\pa_c\f{\6^bf}{|\pa f|} = C^a, 
\label{ctsurface1}\\
&&-\frac{p^cp^d}{m|\pa f|}\,\left[ \pa_c\pa_d\f{\6_b f}{|\pa f|}+3(\pa_c\f{\6^a f}{|\pa f|})(\pa_d\f{\6_af}{|\pa f|})\,\f{\6_bf}{|\pa f|}\right]-2\lambda\,\f{\6^cf}{|\pa f|}\,\pa_c\f{\6_b f}{|\pa f|} = D_b,
\label{ctsurface2}\\
&&p^c(\f{\6_bf}{|\pa f|}\pa_c\f{\6_af}{|\pa f|}-\f{\6_af}{|\pa f|}\pa_c\f{\6_bf}{|\pa f|})= O_{ab},
\label{ctsurface3}
\eea
where $C^a$, $D_b$ and $O_{ab} = -O_{ba}$ are constant numbers. In this case,  (\ref{Projectedcoordinates}) is a system of linear equations with numeric coefficients (from $R$). Therefore, one can find the independeent degrees of freedom as follows.
Since the projected momenta 
$p^\ast_\lambda$ vanishes, we should look for three more vanishing or linearly
dependent degrees of freedom in (\ref{Projectedcoordinates}). They can be 
obtained by noting that not all the coefficients in the r.h.s. of 
(\ref{Projectedcoordinates}) are independent since they are entries of the
projector (\ref{Projector}). Therefore, from the condition
\be
\Lambda^2 = \Lambda
\label{Projectorcondition}
\ee
we derive the following relations
\bq
&&v_a\widetilde\delta\ud ab-k^a\omega_{ab}=v_b\;,\label{Ccoeff1}\\
&&k^a\widetilde\delta\du ab=k^b\;,
\label{Ccoeff2}\\
&&\widetilde\delta\ud ac\,\widetilde\delta\ud cb=
\widetilde\delta\ud ab\;,\label{Ccoeff3}\\
&&\omega_{ac}\widetilde\delta\ud cb+\widetilde\delta\du ac\omega_{cb}=\omega_{ab}\;.
\label{Ccoeff4}
\eq
It is important to note that (\ref{Projectorcondition}) is not a supplementary
condition imposed by hand. Rather, it is a built-in relation in the formalism
because $\Lambda$ is constructed to be the local projector on the constrained
surface. Therefore, the relations (\ref{Ccoeff1})-(\ref{Ccoeff4}) are natural
constraints.  $n^an_b$ is the projector on the normal direction and 
$\widetilde\delta\ud ab$ is the projector onto the tangent space to $\Sigma$.
We can use these projectors to split the coordinates 
(\ref{Projectedcoordinates}) into normal and tangential. Then it is easy to 
show that  
\bea
x^{\ast a}_t &=& \widetilde\delta\ud ab\,x^{\ast b}
=\widetilde\delta\ud abx^b+k^ap_\lambda\;,\label{Xphysical}\\
p^\ast_{ta}&=&\widetilde\delta\du ab\left(\omega_{bc}x^c+
v_bp_\lambda+p_b\right)\;.\label{Pphysical}
\eea
represent the independent degrees of freedom. They are tangential to the
surface $\Sigma$ as they should be. The dependent and null degrees of freedom 
are given by the following relations
\bq
\lambda^\ast &=& v_ax^{\ast a}_t-k^ap^\ast_{ta}\;,\\
x^{\ast a}_n &=& 0\;,\\
p^\ast_\lambda    &=& 0\;,\\
p^\ast_{na}  &=& n_an^b\omega_{bc}x^{\ast c}_t\;.
\label{Dependentrelations}
\eq
The equations above elliminate four degrees of freedom.
Therefore, the number of the
local physical degrees of freedom is $2N-2$ as expected. Note that they are 
expressed in terms of the global indices in the initial $R^N$ space. One can 
pass to a local coordinate system on $\Sigma$ by performing a general 
coordinate transformation from the origin of $R^N$ to the point $P$ on 
$\Sigma$. In order to elliminate the normal direction, one has to pick-up one 
of the directions of the local coordinate system parallel to the normal.
However, the tangential directions are determined only up a $SO(N-1)$ 
transformation.  

The physical Hamiltonian of the system should be entirely written in terms of the 
tangential degrees of freedom and it is given by the folowing relation
\be
H^{\ast} = \frac{1}{2m}p^\ast_{ta}p^{\ast a}_{t} +  
\frac{1}{2m} n^a n^b\omega_{ac}\omega_{bd}x^{\ast c}_tx^{\ast d}_t\ 
\label{PhysicalHamiltonian}.
\ee
The local quantization of the system should be done by using (\ref{PhysicalHamiltonian}).
In order to quantize the system globally on $\Sigma$, additional information
should be given, as for example, the range of the local coordinates and the
topological structure of the theory. It is not clear yet if this information
could be included in a global projector of the surface or if it should appear
from gluing the local operators on neighbour openings.

The two cases analysed above, i. e. when $\Lambda$ containts general functions on $\xi$ and when it contains just constants, represent the two extreme cases of the symplectic projector method. In the first case one cannont provide a general method for finding the independent degrees of freedom, while in the second case they are given by solving a system of linear equations with coefficients from $R$. Between the two cases, there are particular surfaces for which the physical degrees of freedom can be computed from the (\ref{Projectedcoordinates}), but the way of solving this system depends on the specific model. However, for all cases the physical Hamiltonian is given by (\ref{PhysicalHamiltonian}) and it should be used for local quantization.
In order to take into account the global properties of the constraint surface, one should extend this formalism. However, even if one limitates to local analysis, there are surfaces for which the local physical coordinates obtained from the symplectic projector vanish. 

\section{Discussions}

As was already noted, the symplectic projector given by the relation (\ref{Projector}) depends on the coordinates of the phase space $\left( x^a, p_a \right)$. This implies that the splitting of the projected coordinates into independent and dependent coordinates also depends on the position of the point at which this analysis is made on the constrained surface. Moreover, from the relations (\ref{Pphysical}) and (\ref{Dependentrelations}) we see that the variables are really independent only if the entries of the symplectic projector are real numbers. Also, note that the equations (\ref{Pphysical}) and (\ref{Dependentrelations}) imply
that some of the physical coordinates and momenta can vanish on certain surfaces.
This can be seen by taking 
$p_{\l } = 0$ on the surface in (\ref{Pphysical}). For example, the physical coordinates and momenta vanish whenever
\be
\6_a f \6^b f = {\delta}_{a}^{b}\6_c f \6^c f .
\label{zerocoordmomenta}
\ee
Also, only momenta can vanish if
\be
(\6_c\6_a\6_b f - \6_b f \6_a\6_c f )x^c = - {\delta}_{ab}. 
\label{zeromomenta}
\ee

To conclude, we have discussed the possibility of applying the symplectic projector to determine the degrees of freedom of a particle on a surface. From the present analysis results that the independent coordinates and momenta can be found exactly only for certain surfaces. If the components of the projector are constant, then the solution can always be obtained from a linear system of algebraic equations with constant coefficients. 
We note that finding nontrivial solutions to the equations (\ref{ctsurface1}), (\ref{ctsurface2}) and (\ref{ctsurface3}) represents an interesting problem.
Also, there are surfaces for which these relations do not hold but which admitt solutions to the system (\ref{Projectedcoordinates}). In both cases, the local quantization of the system should be done by using (\ref{PhysicalHamiltonian}). 
In the general case, there is no systematic method to split the projected coordinates into a set of independent variables and a linearly dependent one. 
The dynamics of the particle on a plane in $R^3$ can be considered as a simple example of a surface for which the projector is constant. This case is somewhat degenerate since the movement is free and the definition of the plane is a matter of chosing a coordinate system in $R^3$. However, as one can easily check out, the system (\ref{Projectedcoordinates}) gives the correct Hamiltonian and the degrees of freedom. As an example of symplectic projector that depends on the coordinates but whose Hamiltonian can be obtained by applying the symplectic projector method one can take the circle $S^1$. Then the symplectic projector gives the Hamiltonian equal to the kinetic energy in agreement with the Dirac algorithm. This result is also true for $S^n$ and represents one of the intermediate cases discussed previously. These examples illustrate the above discussion.

In order to quantize the system globally on $\Sigma$, additional information
should be given, as for example, the range of the local coordinates and the
topological structure of the theory. It is not clear yet if this information
could be included in a global projector of the surface or if it should appear
from gluing the local operators on neighbour openings. The globality problem 
invites to extending the local symplectic operator method to global theories. We hope to report on these topics in a future paper \cite{New}.

{\bf Acknowledgments}

M. A. S. and I. V. V. would like to thank to J. A. Helayel-Neto for 
hospitality at CBPF during the preparation of this work. M. A. de Andrade acknowledeges to J. A. Helayel-Neto and M. Botta for useful discussions.
I. V. V. was supported by the FAPESP Grant 02/05327-3.

\end{document}